\documentclass[runningheads]{llncs}
\usepackage[numbers,sort&compress]{natbib}

\usepackage{times}
\usepackage{latexsym}

\usepackage[T1]{fontenc}

\usepackage[utf8]{inputenc}

\usepackage{microtype}

\usepackage{amsfonts}
\usepackage{amsmath}
\usepackage{amssymb}

\usepackage{pifont}
\newcommand{\cmark}{\ding{51}}%
\newcommand{\xmark}{\ding{55}}%

\usepackage[table]{xcolor}
\usepackage{graphicx}
\usepackage{booktabs}
\usepackage{multicol}
\usepackage{multirow}
\usepackage{todonotes}

\begin{document}
\title{Parameter-efficient Zero-shot Transfer for Cross-Language Dense Retrieval with Adapters}
\titlerunning{Parameter-efficient ZS Transfer for CL Dense Retrieval with Adapters}

\author{ Eugene Yang\inst{1}\orcidID{0000-0002-0051-1535} \and 
 Suraj Nair\inst{1,2}\orcidID{0000-0003-2283-7672} \and \\
 Dawn Lawrie\inst{1}\orcidID{0000-0001-7347-7086} \and 
 James Mayfield\inst{1}\orcidID{0000-0003-3866-3013} \and
 Douglas W. Oard\inst{1,2}\orcidID{0000-0002-1696-0407}}

 \authorrunning{E. Yang et al.}
 \institute{
 HLTCOE. Johns Hopkins University, Baltimore  MD 21211, USA \\
 \email{\{eugene.yang,lawrie,mayfield\}@jhu.edu}
 \and 
 University of Maryland, College Park MD 20742, USA  \email{\{srnair,oard\}@umd.edu}
 }

\maketitle              %
\begin{abstract}
A popular approach to creating a zero-shot cross-language retrieval model is to substitute a monolingual pretrained language model in the retrieval model with a multilingual pretrained language model such as Multilingual BERT. This multilingual model is fined-tuned to the retrieval task with monolingual data such as English MS MARCO using the same training recipe as the monolingual retrieval model used.
However, such transferred models suffer from mismatches in the languages of the input text during training and inference. 
In this work, we propose transferring monolingual retrieval models using adapters, a parameter-efficient component for a transformer network. 
By adding adapters pretrained on language tasks for a specific language with task-specific adapters, prior work has shown that the adapter-enhanced models perform better than fine-tuning the entire model when transferring across languages in various NLP tasks. 
By constructing dense retrieval models with adapters, we show that models trained with monolingual data are more effective than fine-tuning the entire model when transferring to a Cross Language Information Retrieval (CLIR) setting. 
However, we found that the prior suggestion of replacing the language adapters to match the target language at inference time is suboptimal for dense retrieval models. We provide an in-depth analysis of this discrepancy between other cross-language NLP tasks and CLIR. 

\end{abstract}

\section{Introduction}
Cross-language information retrieval (CLIR), where the query and documents are in different languages, is an important problem in information retrieval because it may be the information a searcher
needs is not available in the language used to query for that information. 
Recently, monolingual neural retrieval models have been applied to CLIR~\cite{ColBERT-X, li2021learning, litschko2021cross, huang2021mixed, C3} with the help of multilingual pretrained language models (mPLMs), such as multilingual BERT~(mBERT~\cite{devlin2018bert}) and XLM-RoBERTa (XLM-R~\cite{conneau2019xlmr}), achieving  state-of-the-art effectiveness in CLIR. 

Training a retrieval model from a pre-trained mPLM with existing monolingual training data, such as MS MARCO~\cite{bajaj2018ms} and Natural Questions~\cite{kwiatkowski-etal-2019-natural}, and transferring the model to a CLIR setting zero-shot is a popular approach. 
Relying on an mPLM for language transfer imposes limitations, however, both because mPLMs must learn over a larger (or less nuanced) vocabulary than a corresponding single-language Pretrained Language Model (PLM) would need and because task-specific training can inadvertently amplify language-specific behaviors when only task-specific learning is desired.
An alternative approach matches the inference-time task more closely at training time by translating either or both of the training queries and the training documents to match the expected language pair~\cite{ColBERT-X, shi2021cross}. However, this can be expensive, and its effectiveness naturally depends on the quality of the machine translation~\cite{ColBERT-X}.  
Both approaches involve compromises that are difficult to manage because they conflate language-specific and task-specific training.  
In this paper, we separate those two training tasks by introducing language-specific pretrained parameters into the neural architecture using an adapter cascade. 

An adapter is a plug-in component for transformer models that ``adapts'' the pretrained model to a specific language and/or task~\cite{houlsby2019adapter, pmlr-v97-stickland19a, bapna-firat-2019-simple}. This approach, first applied for Natural Language Processing (NLP) tasks in 2019, seeks to overcome mismatches
in training data without fine-tuning the entire model. 
Adapters are composable and can be used to create a chain or a stack of adapters, each with different purposes and effects. 
A language adapter followed by a task-specific adapter is a popular chain. It has been shown to be effective when zero-shot transferring a model from one language to another by swapping the language adapter used at training time with a target language adapter~\cite{pfeiffer2020madx}.  
Specifically, an adapter introduces a small number of parameters at each transformer layer that projects the representations over a bottleneck structure with a residual connection~\cite{houlsby2019adapter, pfeiffer2020adapterfusion}. The original PLM is frozen during training, and only the adapter parameters are updated. 
At inference time, the language adapter is replaced with one trained on the target language. Prior work has shown this zero-shot-with-replacement transfer provides better performance than transferring a fine-tuned model~\cite{pfeiffer2020madx}.  

Such language-transferring methods have not been tested with dense retrieval models. We address this gap by constructing a dense retrieval model with a mPLM and adapters. The retrieval model is fine-tuned with the pretrained English adapter attached using English MS MARCO. At inference time (in CLIR, at indexing time and at search time), the text is encoded by the trained model with a language adapter in the target language. 
With appropriate adapter settings, models constructed with 
adapters are, on average, 14\% more effective than those that fine-tune the entire transformer model while updating only 0.5\% as many parameters. %
To our surprise, however, and contrary to prior findings~\cite{pfeiffer2020madx}, replacing the language adapter with the one trained for the target language does \textit{not} lead to better effectiveness in dense retrieval. 
In this paper, we analyze this surprising finding in-depth by testing different queries, adapter sizes, mPLMs, and dense retrieval models.

\begin{figure}[t]
    \centering
    \includegraphics[width=\linewidth]{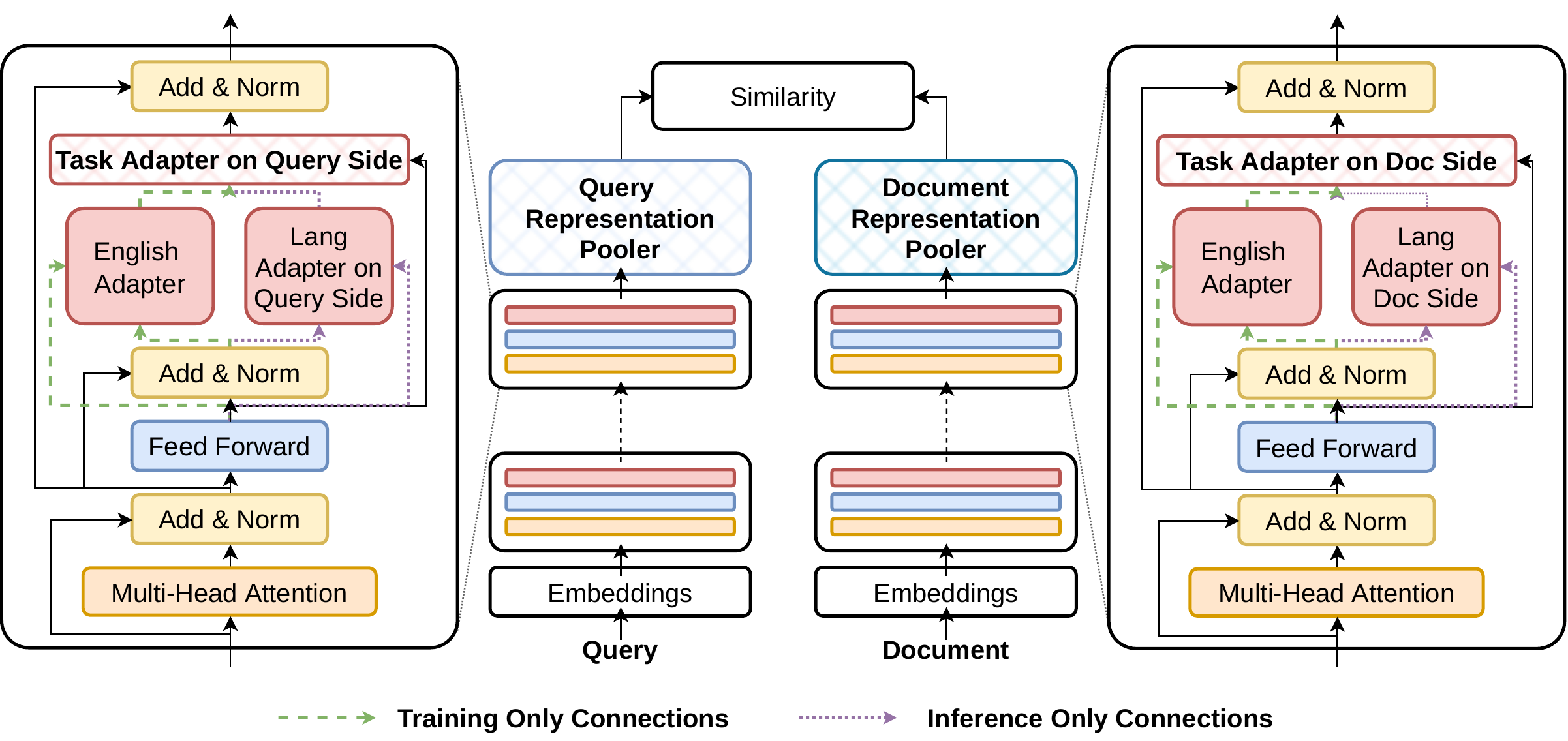}
    \caption{Model Overview. The green dashed connections are used only during training; 
    the purple dotted connections are inference only (including indexing and searching). The components with cross-hatches are the trainable components in our model and can be thought of as the task adapters. }
    \label{fig:model-diagram}
\end{figure}

\section{Background}

One of the main challenges in building CLIR systems is to bridge the language barrier between the query language and the language of the documents.
Prior works have used Machine Translation (MT) systems to translate the queries or the documents, thereby reducing the problem to matching in the monolingual space~\cite{zhou2012translation, nie2010cross, galuvsvcakova2021cross}. Generating dense representations of queries and documents provides an alternative approach in which matching happens in a shared multilingual vector space; this approach is popularly known as {\em dense retrieval}. Pre-BERT dense retrieval models used cross-lingual non-contextualized word representations to perform CLIR matching~\cite{yu2020study, li2018learning, gupta2017continuous}. The advent of large multilingual pretrained language models (mPLMs) such as mBERT and XLM-R enables contextualized representations, which have fueled a wave of dense neural CLIR systems~\cite{shi2021cross, litschko2021cross, li2021learning, ColBERT-X}.

Training a dense neural CLIR system is challenging compared to its monolingual counterpart due to a lack of high-quality training collections. Using off-the-shelf mPLMs directly to do CLIR produces subpar results since the representations are generic and not specific to the downstream CLIR task~\cite{litschko2021cross}. Existing training approaches rely on transfer learning either in a zero-shot~\cite{shi2021cross} or a translate-train~\cite{ColBERT-X} setting using a large relevance-labeled collection such as MS MARCO~\cite{bajaj2018ms, bonifacio2021mmarco}. In the zero-shot setting, an mPLM trained using a high-resource language (e.g., English) is applied directly to the target language. The zero-shot approach thus relies on the generalization capability of mPLMs to tackle the CLIR task. In the translate-train scenario, the monolingual training collection is translated to the target language of interest, and the system is trained on the translated data. Not only is this approach resource intensive, but its effectiveness depends on the quality of translations~\cite{ColBERT-X}. Because of these limitations, this paper focuses on the zero-shot setting for the training and evaluation of CLIR systems.

The most common approach to training a dense retrieval model initialized with a PLM requires updating the weights of all the model parameters.
Adapters~\cite{houlsby2019adapter, pfeiffer2020adapterfusion, bapna-firat-2019-simple, ruckle2020adapterdrop, karimi2021compacter} offer a complementary parameter-efficient approach to training language models for a specific downstream task. The key idea is to insert additional parameters between the transformer layers that can be tailored to capture different modeling aspects, such as language-specific or task-specific modeling~\cite{pfeiffer2020madx, pfeiffer2020unks}. Only these additional language- or task-specific parameters are updated during training while keeping the rest of the parameters frozen. While adapters are increasingly  used in NLP applications~\cite{pfeiffer2020adapterfusion, bapna-firat-2019-simple, ruckle2020multicqa, he-etal-2021-effectiveness, winata2020adapt}, adopting such a parameter-efficient training approach remains underexplored in IR. Recent work explores parameter-efficient training approaches for cross-language neural reranking~\cite{litschko2022parameter} that leverages a cross-encoder architecture, which prevents one from indexing the documents offline. 
In this work, we explore incorporating adapters into dense retrieval models, which independently encode queries and documents. 

We chose two dense CLIR retrieval models in our experiments as representatives of dense CLIR approaches. The first is 
Dense Passage Retrieval (DPR)~\cite{DPR}, which 
computes the inner products between classification (CLS) tokens for the query and each document. The second is  ColBERT-X~\cite{ColBERT-X}, a CLIR model that extends ColBERT~\cite{ColBERT}.  ColBERT-X combines three key ideas.  Drawing insight from BERT~\cite{devlin2018bert}, it represents documents using contextualized embeddings, with the embedding for each term instance influenced by that instance's context.  Contextual embeddings better represent meaning than simple term occurrence. Leveraging both multilinguality and improved pre-training from XLM-R~\cite{conneau2019xlmr}, ColBERT-X seeks to generate similar contextual embeddings for terms with similar meaning, regardless of their language.  Drawing its structure from ColBERT, ColBERT-X limits ranking latency by separating query and document transformer networks to support offline indexing.  ColBERT scores documents by focusing query term attention on the most similar contextual embedding in each document. 
In the rest of the paper, we use ``ColBERT'' when referring to the neural retrieval architecture and ``ColBERT-X'' specifically for the CLIR model.

\section{Dense CLIR Model with Adapters}

In this section, we introduce our retrieval model architecture constructed using adapters. 
Figure~\ref{fig:model-diagram} illustrates the overview of the model. 

The queries and documents are passed through separate adapter-transformer encoders constructed over
a shared, frozen pretrained transformer model with two different adapter settings. 
The hidden states of the final adapter-transformer layer are passed through a pooler to form a smaller representation~\cite{DPR, ColBERT}. 
Notice that there is a language adapter in the query representation
stack and the document representation stack. This means that for CLIR, a different language adapter can be used to encode the queries
and the documents.
Finally, the score of a document given a query is produced by calculating the similarity between the query and document representations. 
This general adapter architecture can be applied to token-level matching models, such as ColBERT, or to passage-level models, such as DPR, by using appropriate similarity functions. 

\subsection{Language and Task Adapters}

Language and task adapters attached to the frozen mPLM share the same architecture but record different knowledge. 
As proposed by \citet{houlsby2019adapter} and refined by \citet{pfeiffer2020adapterfusion}, an adapter consists of a down-projection that shrinks the normalized hidden state to a \textit{bottleneck} vector, followed by an up-projection that restores the size of the vector to allow the combination with the original hidden state. Let the $\mathbf{D}\in \mathbb{R}^{h \times (h/r)}$ and $\mathbf{U}_l\in \mathbb{R}_l^{(h/r) \times h}$ be the down- and up-projections of layer $l$ where $h$ is the size of the hidden state in the transformers, and $r$ is the reduction factor. An adapter of layer $l$ can be expressed as 
\begin{equation}
    \mathbf{A}_l(\mathbf{h}_l, \mathbf{r}_l) = \mathbf{U}_l( \text{ReLU}( \mathbf{D}_l(\mathbf{h}_l) ) ) + \mathbf{r}_l. 
    \label{eq:adapter}
\end{equation}
where $\mathbf{h}_l$ and $\mathbf{r}_l$ are the hidden state and the residual of layer $l$. 

While there are alternative architectures for constructing adapters, \citet{pfeiffer2020adapterfusion} has shown that the form expressed in Equation~\ref{eq:adapter}, usually referred to  as the \textit{Pfeiffer adapter}, is more effective than others, including the original adapter architecture proposed by \citet{houlsby2019adapter} (the \textit{Houlsby adapter}), which requires two bottleneck structures for each adapter in a layer. 
In this study, we use Pfeiffer adapters due to their simplicity and the wider availability of pretrained Pfeiffer adapters.\footnote{However, the robustness and generalizability of Pfeiffer adapters has recently been questioned; see \url{https://github.com/Adapter-Hub/adapter-transformers/issues/168}.} 

In each layer, we add a trainable retrieval adapter as our task adapter for the encoders. The retrieval adapters on the query and document side $\mathbf{QA}_l$ and $\mathbf{DA}_l$ take the output of the language adapter $\mathbf{LA}^{Q}_l$ and $\mathbf{LA}^{D}_l$  on the query and document side, respectively, to form the final language- and retrieval-aware hidden states. The down- and up-projections are not shared across adapters.  Specifically, the chained adapters can be expressed as
\begin{align*}
    f^{Q}_l( L_Q, \mathbf{h}_l, \mathbf{r}_l ) &= \mathbf{QA}_l\left(
        \mathbf{LA}^{Q}_l(\mathbf{h}_l, \mathbf{r}_l), \mathbf{r}_l
    \right) \\
    f^{D}_l( L_D, \mathbf{h}_l, \mathbf{r}_l ) &= \mathbf{DA}_l\left(
        \mathbf{LA}^{D}_l(\mathbf{h}_l, \mathbf{r}_l), \mathbf{r}_l
    \right)
\end{align*}
where the adapters $\mathbf{LA}^{Q}$ and $\mathbf{LA}^{D}$ can be language adapters pretrained on any language.

\subsection{Training and Inference Adapter Settings}

During retrieval fine-tuning, an English adapter is used as the language adapter on both the query and document sides to match the native language of MS MARCO~\cite{bajaj2018ms}, which is the common training resource used to train neural retrieval models. The mPLM and English adapters are frozen for both query and document encoders. The trainable components are the retrieval adapters, $\mathbf{QA}_l$ and $\mathbf{DA}_l$ for all layers $l$, and the poolers (illustrated with cross-hatch background in Figure~\ref{fig:model-diagram}). The number of parameters being updated in the model is around 0.5\% to 10\% depending on the size of the adapters, which is set by the reduction factor $r$. 

During inference, including both indexing and searching, the language adapters are set to match the language of the input text.

\section{Experiment Setup}

\subsection{Data}
\begin{table}[t]
\setlength\tabcolsep{0.55em}

    \caption{Collection statistics.}
    \label{tab:collection-stats}
\centering
\begin{tabular}{l|rrrr|rr}
\toprule
\multirow{2}{*}{Collection}
              &     \multicolumn{4}{c|}{CLEF 2003} & \multicolumn{2}{c}{HC4} \\
              &     German &  Finnish &   French &  Russian &    Persian &    Chinese  \\
\midrule
\# documents  &    294,805 &   55,344 &  129,804 &   16,715 &    486,486 &    646,305 \\
\# passages   &  1,642,725 &  378,108 &  762,578 &  105,534 &  3,098,104 &  3,645,078 \\
\# queries    &         60 &       60 &       60 &       62 &         50 &         50 \\
\bottomrule
\end{tabular}
\end{table}
\begin{table}[tb]
\setlength\tabcolsep{0.55em}

    \caption{Relative model size in percentage of parameters of each type of adapter. The number of parameters of the task adapters, i.e. query and document adapters, for DPR is two times the numbers presented in the table given the lack of sharing between  query and document adapters. }
    \label{tab:model-param-percentage}
\centering
\begin{tabular}{l|cccc}
\toprule
            & Full Model 
            & Lang. Adapter 
            & Task Adapter ($r=16$) 
            & Task Adapter ($r=2$) \\
\midrule
mBERT       &  100\% &  3.987\%  &  0.503\%  &  3.987\% \\
XLMR        &  100\% &  2.551\%  &  0.322\%  &  2.551\% \\
\bottomrule
\end{tabular}

\end{table}

We use MS MARCO v1 triples~\cite{bajaj2018ms} as our training data.
Texts are tokenized with the tokenizer required by the language model, using a fixed length of 32 tokens for queries and 180 tokens for documents by padding or truncation. 

For evaluation, we use the HC4~\cite{lawrie2022hc4} collection for Chinese and Persian, which consists of newswire documents from Common Crawl; and from the Cross-Language Evaluation Forum (CLEF) 2003 collections~\cite{Braschler2003-zs} for German, Finnish, French, and Russian (which also consist of news articles).
Collection information is summarized in Table~\ref{tab:collection-stats}. 

We use English title queries for our primary evaluation condition to simulate web search queries.  
As a contrastive condition, we also evaluate the search using human-translated titles;
this models a monolingual retrieval problem in the document language as an alternative to the CLIR zero-shot transfer scenario.
Documents are tokenized and split into overlapping passages of 180 tokens with a sliding window of 90 tokens. We apply the MaxP~\cite{Dai2019-sl} technique to use the maximum score across a document's passages as the document's score. 
Note that texts in every language are segmented only by the mPLM's tokenizer, which may result in over-segmentation for Chinese. 

\subsection{Models and Training Settings}

\subsubsection{Retrieval Models.}
We experiment with both DPR and ColBERT models by using adapters to construct the model.
To match the original design of DPR, the retrieval adapters on the query and document sides, including the poolers, are separate models with independent trainable parameters; to construct ColBERT, the encoders are shared across queries and documents, i.e., a single retrieval adapter sharing the same set of parameters, as do the poolers. The pooler performs a linear transformation on the hidden state of the CLS token from the last layer for DPR; for ColBERT, the pooler projects the final hidden states of each token to vectors of size 128. The reduction factors of the query and document adapters are 16, which is suggested by our pilot study. We also provide evidence in the next section that larger adapters are not more effective for CLIR. 

The language adapters are pretrained on Wikipedia articles with masked language modeling for 250k steps and a batch size of 64~\cite{pfeiffer2020madx}, which we acquired from AdapterHub. The reduction factors of these language adapters are fixed to 2 by \citet{adapterhub}. 

During training, the training triples are encoded by the corresponding encoders, and a binary softmax loss is formed by the similarities between the query and the two passages. We use Adam~\cite{kingma2014adam} as the optimizer. 

\subsubsection{Pretrained Language Models.}
We also test the retrieval models with two mPLMs: Multilingual BERT-Base-Cased (mBERT) and XLM-RoBERTa-Base (XLM-R). Based on our pilot studies and suggestions from prior work~\cite{ColBERT-X, C3}, we set the learning rates to $1\times 10^{-5}$ and $5\times 10^{-6}$ for mBERT and XLM-R, respectively. Due to the limited availability of language adapters for XLM-R on AdapterHub, we only evaluate German, Russian, and Chinese on XLM-R and all six languages for mBERT. 

For completeness, we also construct a pair of retrieval models with mBERT without adding the language adapters. In these ablated DPR and ColBERT models, only retrieval adapters are added to the architecture, relying on the mPLM to encode text in different languages and thus offering insight into the effect of the language adapters.

\subsubsection{Evaluation.}
We compare the adapter models by fine-tuning the entire transformer model also with English MS MARCO v1~\cite{bajaj2018ms} and evaluating by retrieving non-English documents based on English queries. This has been referred as a \textit{zero-shot} transfer approach in prior work~\cite{shi2021cross, litschko2021cross, ColBERT-X, C3}. As summarized in Table~\ref{tab:model-param-percentage}, adapters update only up to 5\% of the parameters compared to full model fine-tuning during training.

We report retrieval effectiveness using nDCG@100. However, the conclusions we draw from nDCG@100 are similar to what we see with MAP and nDCG@10. 
\section{Results and Analysis}

\begin{table}[t]
\setlength\tabcolsep{0.45em}
\renewcommand{\arraystretch}{1}
\renewcommand{\b}[1]{\textbf{#1}}

\newcommand{\s}{$\dagger$}
\newcommand{\n}{\phantom{$\dagger$}}

\renewcommand{\v}{$^*$}
\newcommand{\x}{\phantom{$^*$}}

\newcommand{\z}{\phantom{1}}

\caption{
nDCG@100 of on six languages using mBERT with English queries. 
The reduction factors of retrieval adapters are 16. 
$\mathbf{LA}^Q$ and $\mathbf{LA}^D$ are the language adapters used at inference time; E and D indicate English and the language documents' native language respectively. FMFT is Full-Model Fine-Tuning.
Cells with \s $ $ are statistically significantly \textbf{different} (can be better or worse) with 95\% confidence from the full model fine-tuning for the corresponding retrieval model using a paired $t$-test with Bonferroni correction for six languages; \v $ $ indicates significant differences between the E-E and E-D rows under the same testing criteria. Significance tests for the Avg. column are not corrected for multiple tests.
}\label{tab:main-results}

\centering
\resizebox{\columnwidth}{!}{
\begin{tabular}{l|cc|rrrrrr|r}
\toprule
Model & $\mathbf{LA}^Q$ & $\mathbf{LA}^D$ & 
German & Finnish & French & Russian & Persian\x & Chinese & Avg.\n\x \phantom{(+ 1\%)} \\
\arrayrulecolor{black}\midrule
\multirow{4.5}{*}{DPR} 
& \multicolumn{2}{c|}{FMFT}
        &    0.323\n &     0.262\n &     0.379\n &     0.363\n &   0.126\n\x &   0.231\n &  0.281\n\x \phantom{(+00\%)}\\
\arrayrulecolor{black!30}\cmidrule{2-10}
& E & E &    0.325\v &     0.319\v &     0.433\v &     0.306\v &   0.197\s\v &   0.235\n &  0.302\v\n\phantom{1} (+7\%)  \\
& E & D &    0.097\s &     0.153\s &     0.218\s &     0.165\s &   0.022\s\x &   0.232\n &  0.148\s\x\phantom{-} (-47\%)  \\
& D & D &    0.118\s &     0.130\s &     0.226\s &     0.182\s &   0.000\s\x &   0.220\s &  0.146\s\x\phantom{-} (-48\%)  \\
\arrayrulecolor{black}\midrule
\multirow{4.5}{*}{ColBERT}
& \multicolumn{2}{c|}{FMFT} 
        &    0.365\n &     0.290\n &     0.460\n &     0.334\n &   0.230\n\x &   0.313\n &  0.332\s\x \phantom{(+00\%)} \\
\arrayrulecolor{black!30}\cmidrule{2-10}
& E & E &    0.400\v &     0.373\n &     0.487\v &     0.327\n &   0.299\n\x &   0.387\s &  0.379\s\v (+14\%) \\
& E & D &    0.235\s &     0.324\n &     0.358\s &     0.280\n &   0.284\n\x &   0.374\n &  0.309\s\x \hspace*{0.5em} (-7\%)\\
& D & D &    0.266\s &     0.324\n &     0.385\n &     0.330\n &   0.238\n\x &   0.366\n &  0.318\s\x \hspace*{0.03em} (-16\%)\\
\arrayrulecolor{black}\bottomrule
\end{tabular}
}
\end{table}

As Table~\ref{tab:main-results} shows, using English language adapters on queries and documents may actually improve nDCG@100, compared to zero-shot full-model fine-tuning (FMFT).  Few of the differences are statistically significant, but on average, the DPR models improve by 7\% and ColBERT by 14\% (compare the FMFT and E-E lines). Applying English language adapters on both the query and document sides thus provides a parameter-efficient approach to zero-shot training for a CLIR model, updating only 1\% (for DPR) or 0.3\% (for ColBERT) of the parameters that would require updates in full-model fine-tuning (ColBERT adapters have even fewer parameters to train because they share query and document encoders). 

Surprisingly, however, using a language adapter that matches the document language on the document side of the encoder stack at inference time results in a statistically significant 47\% average degradation from FMFT for DPR, and a (sometimes significant) 7\% average degradation for ColBERT when compared to ColBERT FMFT. 
To compare the two inference adapter settings, since the representations of the queries are the same (E-E and E-D), 
our empirical evidence suggests that the content representations with the language adapter tuned for English better match the query representations from that same language adapter than do content representations produced by a language adapter trained for the document language. As the degradation is across all languages and both retrieval models, we believe it is not due to the quality of an individual pretrained language adapter but rather it is due to what it models. 
Since the language adapters are trained independently on Wikipedia articles, the hidden states produced by the language adapters are not language-agnostic. 
Therefore, changing the adapter setting at inference time creates a mismatch in the hidden representations that the downstream retrieval adapters do not anticipate, resulting in worse retrieval effectiveness. 

That result stands in contrast with \citet{pfeiffer2020madx}, which found the primary advantage of adapters in language transfer to be the ability to ``adapt'' the model to focus on the target language. To the extent such benefits exist, DPR and ColBERT fail to exploit them effectively.  
Furthermore, working with two languages is a unique challenge in cross-language dense retrieval compared to other cross-language transfer learning tasks, where the task is monolingual. 
The experimental setting in \citet{pfeiffer2020madx} transfers an English model to a task, such as NER or answer generation in question-answering, %
solely in another language. 
As a result, such transfers do not require any properties \textit{between} language adapters but only properties within the language. 
Dense retrieval is different -- calculating similarities between the representation of the queries in one language and documents in another requires inter-language-adapter similarities. 
So crossing the language barrier with language adapters appears to be worse than relying on multilingual language models, which are trained jointly with text in multiple languages, even though no alignment information is used in either training process. 

From the apparent (but rarely significant) improvement of E-E over FMFT, however, we might also conclude that what we have called an ``English'' language adapter is perhaps actually better thought of as a ``zero-shot'' language adapter;
our results suggest that it does not harm, and may actually improve the mPLM's representation, even across languages.

Unsurprisingly, when the language adapters at both the query and the document sides are replaced  with adapters trained in the document language (the D-D condition), nDCG@100 is on par with the E-D condition.

We further explore the impact of language adapters in the next section by omitting both language adapters and using only the retrieval adapters on the query and the document sides.

\subsection{Monolingual Retrieval with Translated Queries}
\begin{table}[t]
\setlength\tabcolsep{0.45em}
\renewcommand{\b}[1]{\textbf{#1}}
\newcommand{\s}{$\dagger$}
\newcommand{\n}{\phantom{$\dagger$}}
\renewcommand{\v}{$^*$}
\newcommand{\x}{\phantom{$^*$}}
\newcommand{\z}{\phantom{1}}

\caption{nDCG@100 on six languages using mBERT with human translated queries. All adapter models use language adapters followed by retrieval adapters ($r=16$) in the encoder. \s $ $ indicates statistically significant differences between the fully fine-tuned model and the corresponding adapter model; \v $ $ indicates the statistically better model between E-E and D-D.  Testing approach is identical to Table~\ref{tab:main-results}.}\label{tab:qt-results}

\centering
\resizebox{\columnwidth}{!}{
\begin{tabular}{l|cc|rrrrrr|r}
\toprule
Model & $\mathbf{LA}^Q$ & $\mathbf{LA}^D$ & 
German & Finnish & French & Russian & Persian\x & Chinese & Avg. \phantom{(+ 1\%)} \\
\arrayrulecolor{black}\midrule
\multirow{3.5}{*}{DPR}
& \multicolumn{2}{c|}{FMFT}
        &    0.242\n &     0.246\n &     0.443\n &     0.425\n &   0.199\n\x &   0.442\n &  0.333\n \phantom{(+ 1\%)} \\
\arrayrulecolor{black!30}\cmidrule{2-10}
& E & E &    0.215\v &     0.278\n &     0.438\v &     0.387\v &   0.307\s\v &   0.447\v &  0.345\v \phantom{-}(+3\%) \\
& D & D &    0.119\s &     0.233\n &     0.292\s &     0.287\s &   0.013\s\x &   0.446\n &  0.232\s (-30\%)\\
\arrayrulecolor{black}\midrule
\multirow{3.5}{*}{ColBERT}
& \multicolumn{2}{c|}{FMFT}
        &    0.372\n &     0.536\n &     0.503\n &     0.511\n &   0.354\n\x &   0.531\n &  0.468\n \phantom{(+ 1\%)} \\
\arrayrulecolor{black!30}\cmidrule{2-10}
& E & E &    0.353\v &     0.446\s &     0.565\v &     0.515\n &   0.412\n\x &   0.531\n &  0.470\n (+.4\%) \\
& D & D &    0.305\s &     0.609\n &     0.510\n &     0.496\n &   0.410\n\x &   0.545\n &  0.479\s \hspace*{0.1em} (+2\%) \\
\arrayrulecolor{black}\bottomrule
\end{tabular}
}

\end{table}

To evaluate adapter models in a monolingual setting as was done in \citet{pfeiffer2020madx}, we use manually-translated queries in the document language to form a non-English monolingual retrieval; results are summarized in Table~\ref{tab:qt-results}. 
Note that monolingual retrieval where documents are retrieved with queries in the same language is an easier problem for a model trained with English data. 
Both DPR and ColBERT perform better in almost all languages with a translated query (comparing across Table~\ref{tab:main-results} and \ref{tab:qt-results}), with ColBERT benefiting more from token representations in the same language. 

When using language adapters to match input text (E-D line in Table~\ref{tab:main-results} and D-D line in Table~\ref{tab:qt-results}), the DPR adapter model improves 56\% from 0.148 to 0.232 compared to 19\% (0.281 to 0.333) for the FMFT DPR model;
this implies that the adapter model benefits more from matching representations of text in the same language. 
This improvement difference further confirms our observation in the previous section that matching representations across language adapters in CLIR poses a unique challenge that is not transferred to monolingual tasks in different languages.  

Likewise, the ColBERT adapter model improves 55\% from 0.309 to 0.479 when querying with translated queries compared to 40\% (0.332 to 0.468) for FMFT. This also suggests that matching representations produced by the same language adapter is more favorable for retrieval models constructed by adapters. 
Furthermore, ColBERT with both language adapters replaced at inference time performs slightly better than the FMFT, despite the difference not being statistically significant. This result suggests that the quality of the token representations is at least on par with the original language model when compared within a specific language. 

On the other hand, keeping the English adapters both on the query and the document side in the DPR model at inference time still outperforms the case where both match the input text, with Persian being particularly ineffective.
This implies that changing the model during inference results in less effective sequence representations, even when transferring from one monolingual task to another monolingual one. 

Our results may seem contradictory to the findings of cross-language reranking with adapters presented by \citet{litschko2022parameter}, which concluded that replacing language adapters to match the document language provides the best effectiveness. 
However, since the reranking model evaluated in their work is a cross-encoder, their retrieval adapters are more capable of reconciling the differences between the language adapters at each level with greater modeling power. 
In dense retrieval, the interaction between the two encoders only happens after the text is encoded; this requires the representation, including the hidden states, to be similar to text that conveys similar meaning regardless of the language. 
This similarity property is generally more difficult to obtain in sequence than token representations, which leads to the effectiveness discrepancy we observe here between DPR and ColBERT.

The degree of impact in effectiveness caused by replacing language adapters differs among languages. 
The Chinese adapter perhaps has the best sequence representation quality compared to other languages, with only a 0.001 difference in nDCG@100 for DPR. 
In contrast, the Persian adapter exhibits a 95\% degradation from 0.307 to 0.013 in DPR. However, the difference in Persian when using ColBERT is less severe, suggesting that the quality issue might lie mainly in the sequence representations instead of the token representations. 
This difference implies that language adapters are \textit{still} built differently, and the curse of multilinguality~\cite{conneau2019xlmr} still exists with applying language adapters pretrained for a specific language.

\section{Ablation Studies}

\subsection{Size of Retrieval Adapters}
\begin{table}[t]
\setlength\tabcolsep{0.45em}
\renewcommand{\arraystretch}{1}
\renewcommand{\b}[1]{\textbf{#1}}
\newcommand{\s}{$\dagger$}
\newcommand{\n}{\phantom{$\dagger$}}
\newcommand{\z}{\phantom{1}}

\caption{
nDCG@100 of DPR with large retrieval adapters ($r=2$) on six languages using mBERT with English queries. 
The improvement and the statistical tests are compared against full model fine-tuning results presented in Table~\ref{tab:main-results} with the same testing criteria.
}\label{tab:large-adapter-results}

\centering
\resizebox{\columnwidth}{!}{
\begin{tabular}{cc|rrrrrr|r}
\toprule
$\mathbf{LA}^Q$ & $\mathbf{LA}^D$ & 
German & Finnish & French & Russian & Persian & Chinese & Avg. \phantom{(+ 1\%)} \\
\arrayrulecolor{black}\midrule
E & E &    0.317\n &     0.235\n &     0.393\n &     0.310\n &   0.171\n &   0.230\n &  0.276\phantom{-0} (-2\%)  \\
E & D &    0.115\s &     0.150\s &     0.211\s &     0.213\s &   0.009\s &   0.212\n &  0.152\phantom{-} (-46\%)  \\
\arrayrulecolor{black}\bottomrule
\end{tabular}
}
\vspace{-1em}
\end{table}

To investigate the impact of the size of the adapters in dense retrieval, we train a DPR adapter model using mBERT with retrieval adapters by setting the reduction factor to 2; this reduces the number of hidden states from 768 only to 384. 
Summarized in Table~\ref{tab:large-adapter-results}, the model performs generally similarly to the smaller adapters in both adapter settings (presented in Table~\ref{tab:main-results}) with different variations in the languages. 
The small number of additional parameters introduced by the small adapters is already sufficient to encode knowledge related to the retrieval task. 

The size of the retrieval adapter does not affect the transferability of the model across languages. 
One would expect retrieval adapters with larger capacities to be less vulnerable to changes in upstream language adapters at inference time. However, our experimental results suggest the contrary -- the chaining topology of the adapters prevents the adapter from being robust to the input hidden states regardless of model size. 

Note that our results do not exclude the possibility that our training process is ineffective. Perhaps a more effective retrieval training process, such as hard negative mining~\cite{kalantidis2020hard}, could further improve the model and show a difference with larger adapters.

\subsection{Different mPLM}
\begin{table}[t]
\setlength\tabcolsep{0.45em}
\caption{nDCG@100 on three languages using XLM-R with English queries. The average numbers are only across three languages instead of six in the mBERT results. 
The difference between full model fine-tuning, E-E, and E-D are not significant under a paired $t$-test with Bonferroni correction for three languages. 
}\label{tab:xlmr-results}

\centering
\begin{tabular}{l|cc|rrr|r}
\toprule
Model & $\mathbf{LA}^Q$ & $\mathbf{LA}^D$ & 
German & Russian & Chinese &  Avg. \phantom{(+ 1\%)} \\
\arrayrulecolor{black}\midrule
\multirow{3.5}{*}{DPR} &
\multicolumn{2}{c|}{FMFT}
&            0.328 &     0.369 &   0.213 &  0.303 \phantom{(+ 1\%)} \\
\arrayrulecolor{black!30}\cmidrule{2-7}
& E & E &    0.300 &     0.353 &   0.170 &  0.274 (-10\%)\\
& E & D &    0.289 &     0.329 &   0.163 &  0.260 (-14\%)\\
\arrayrulecolor{black}\midrule
\multirow{3.5}{*}{ColBERT} &
\multicolumn{2}{c|}{FMFT}
&            0.449 &     0.409 &   0.296 &  0.385 \phantom{(+ 1\%)} \\
\arrayrulecolor{black!30}\cmidrule{2-7}
& E & E &    0.442 &     0.423 &   0.260 &  0.375\phantom{1} (-3\%)\\
& E & D &    0.440 &     0.393 &   0.262 &  0.365\phantom{1} (-5\%)\\
\arrayrulecolor{black}\bottomrule
\end{tabular}

\vspace{-1em}
\end{table}

To demonstrate the robustness of the adapter framework across mPLMs, we train a pair of DPR and ColBERT models with XLM-R as the underlying language model and summarize the results in Table~\ref{tab:xlmr-results}. Due to the limited availability of pretrained language adapters for XLM-R, we conducted CLIR experiments on only German, Russian, and Chinese. 
As shown in prior work~\cite{ColBERT-X}, XLM-R is generally a better language model for dense retrieval, which is also suggested by our result. 

Relative to XLM-R FMFT, on average, DPR and ColBERT constructed with adapters provide 90\% and 97\% nDCG@100 by updating only 0.6\% and 0.3\% of the parameters. Those nDCG@100 differences are not statistically significant. This means that we cannot reject the null hypothesis that no difference in nDCG@100 exists. 
Notably, DPR with adapters demonstrates a larger degradation in nDCG@100 than ColBERT, suggesting, again, that sequence representations produced by the adapters may be less effective in crossing the language barrier compared to the token representations. This matches the conclusion drawn from the mBERT results. 

When replacing the language adapter on the document side with the document language at inference time, retrieval effectiveness also drops, albeit less severely than for mBERT variants; this suggests that the hidden states produced by the XLM-R language adapters 
are more similar to what the XLM-R English adapter would be providing to the downstream retrieval adapter.
Despite only comparing with three languages, language adapters trained for XLM-R seem more language-agnostic than language adapters trained for mBERT. 

\subsection{Only Retrieval Adapters}
\begin{table}[t]
\setlength\tabcolsep{0.50em}
\renewcommand{\arraystretch}{1}
\renewcommand{\b}[1]{\textbf{#1}}
\newcommand{\s}{$^*$}
\newcommand{\n}{\phantom{$^*$}}
\newcommand{\z}{\phantom{1}}

\caption{nDCG@100 on six languages using mBERT with and without language adapters. 
The results with language adapters do not replace the language adapters at inference time. Monolingual queries are in the document language.
\s $ $ indicates a statistically significantly better adapter setting with 95\% confidence under the same retrieval model and query language. The tests are done by paired $t$-tests with Bonferroni corrections for 12 tests (six languages and two types of queries).
Significance tests on the Avg. column are not corrected for multiple tests.
}\label{tab:nolang-results}

\centering
\resizebox{\columnwidth}{!}{
\begin{tabular}{l|lc|rrrrrr|r}
\toprule
Model & Query & LA &
German & Finnish & French & Russian & Persian & Chinese & Avg.\n \\
\arrayrulecolor{black}\midrule
\multirow{4.5}{*}{DPR}
& \multirow{2}{*}{English}
 & \xmark &  0.290\n &     0.230\n &     0.356\n &     0.289\n &   0.163\n &   0.220\n &  0.258\n \\
&& \cmark &  0.325\n &     0.319\s &     0.433\s &     0.306\n &   0.197\n &   0.235\s &  0.302\s \\
\arrayrulecolor{black!30}\cmidrule{2-10}
& \multirow{2}{*}{Monolingual}
 & \xmark &  0.193\n &     0.168\n &     0.340\n &     0.397\n &   0.234\n &   0.421\n &  0.292\n \\
&& \cmark &  0.215\n &     0.278\s &     0.438\s &     0.387\n &   0.307\n &   0.447\n &  0.345\s \\
\arrayrulecolor{black}\midrule
\multirow{4.5}{*}{ColBERT}
& \multirow{2}{*}{English}
 & \xmark &  0.405\n &     0.388\n &     0.488\n &     0.319\n &   0.304\n &   0.377\n &  0.380\n \\
&& \cmark &  0.400\n &     0.373\n &     0.487\n &     0.327\n &   0.299\n &   0.387\n &  0.379\n \\
\arrayrulecolor{black!30}\cmidrule{2-10}
& \multirow{2}{*}{Monolingual}
 & \xmark &  0.398\s &     0.556\s &     0.560\n &     0.530\n &   0.438\n &   0.556\n &  0.506\n \\
&& \cmark &  0.353\n &     0.446\n &     0.565\n &     0.515\n &   0.412\n &   0.531\n &  0.470\n \\
\arrayrulecolor{black}\bottomrule
\end{tabular}
}

\end{table}

Finally, we examine models constructed without language adapters, using only retrieval adapters. 
At inference time, the model stays identical to training time and performs zero-shot transfer like the FMFT models. We compare this retrieval-adapter-only model with models keeping the English adapter as the language adapter on both sides during inference. 
 
As summarized in Table~\ref{tab:nolang-results}, English adapters are helpful only when retrieving using DPR.  
For DPR, the final query and document representations may benefit from the additional language model training recorded in the English adapter. 
However, for ColBERT, which operates on token-level representations, English adapters perform on par or worse compared to not using any language adapter. This implies that the English adapter may not drastically improve the multilinguality of the pretrained language model. The original mPLM is already as good as using adapters for transferring knowledge across languages for ColBERT.

\section{Summary and Future Work}

In this work, we investigated dense retrieval models constructed by adapters for CLIR. 
We demonstrated a 7\% and 14\% improvement in nDCG@100 for DPR and ColBERT model by updating only 1\% and 0.5\% of the parameters, respectively, during retrieval fine-tuning. 
Contrary to prior findings, replacing the language adapters to match the language of the input text at inference time does not lead to more effective zero-shot transferred CLIR dense retrieval models. This is especially for DPR, which relies on matching the sequence representations. 
Based on our results and analysis, we conclude that matching representations across languages in CLIR poses a unique challenge to adapters. 
Such issues are less severe when matching token representations, but language adapters are not necessarily more effective than using the mPLM directly. 

There are still a host of questions that should be investigated for adapters in CLIR, including the quality of individual adapters. 
Building language adapters tailored to dense retrieval could further improve the effectiveness of adapter models for the task. 
Developing modeling and indexing techniques that exploit the composability of the adapters could lead to retrieval models that are not just parameter-efficient during training, but also space-efficient.

\bibliographystyle{splncs04nat}
\bibliography{bibio}
\end{document}